\input harvmac
\input epsf

\def\al{\alpha}
\def\at{\tilde{\alpha}}
\def\ap{\alpha'}

\def\p{\partial}

\def\half{{1\over 2}}


\Title{}{\vbox{\centerline{Winding String Condensation}
  \smallskip
  \centerline {and}
\smallskip
  \centerline {Noncommutative Deformation of Spacelike Singularity}  }}

 \centerline{Jian-Huang
She\footnote{}{Emails: jhshe@itp.ac.cn}}

\medskip
\centerline{\it  Institute of Theoretical Physics, Chinese Academy of Science, }
 \centerline{\it P.O.Box 2735, Beijing 100080, P.R. China}
\medskip

\medskip
\centerline{\it  Graduate School of the Chinese Academy of Sciences, Beijing 100080, P.R. China}

\medskip

In a previous paper (hep-th/0509067) using matrix model, we showed
that closed string tachyons can resolve spacelike singularity in
one particular class of Misner space (with anti-periodic boundary
conditions for fermions around the spatial circle). In this note,
we show that for Misner space without closed string tachyons,
there also exists a mechanism to resolve the singularity in the
context of the matrix model, namely cosmological winding string
production. We show that here space and time also become
noncommutative due to these winding strings. Employing optical
theorem, we study the bulk boundary coupling by calculating the
four-open-string cylinder amplitudes.

\Date{Feb. 2006}

\nref\APS{A. Adams, J. Polchinski, E. Silverstein, "Don't panic!
Closed string tachyons in ALE space-times", JHEP 0110 (2001) 029,
hep-th/0108075.}

\nref\sm{John McGreevy, Eva Silverstein, "The Tachyon at the End
of the Universe",hep-th/0506130.}

\nref\horo{Gary T. Horowitz, "Tachyon Condensation and Black
Strings", JHEP 0508 (2005) 091, hep-th/0506166.}

\nref\she{Jian-Huang She, "A Matrix Model for Misner Universe and
Closed String Tachyons", JHEP 0601(2006)002, hep-th/0509067.}

\nref\tai{Yasuaki Hikida, Ta-Sheng Tai, "D-instantons and Closed
String Tachyons in Misner Space", hep-th/0510129.}

\nref\sil{A. Adams, X. Liu, J. McGreevy, A. Saltman, E.
Silverstein, "Things Fall Apart: Topology Change from Winding
Tachyons", hep-th/0502021.}

\nref\hellerman{Simeon Hellerman, "On the Landscape of Superstring
Theory in D > 10", hep-th/0405041; Simeon Hellerman, Xiao Liu,
"Dynamical Dimension Change in Supercritical String Theory",
hep-th/0409071.}

\nref\rsimon{Simon F. Ross, "Winding tachyons in asymptotically
supersymmetric black strings", JHEP 0510 (2005) 112,
hep-th/0509066.}

\nref\newpaper{Haitang Yang, Barton Zwiebach, "Rolling Closed String
Tachyons and the Big Crunch", hep-th/0506076;  Haitang Yang,
Barton Zwiebach, "A Closed String Tachyon Vacuum ?",
hep-th/0506077; Takao Suyama, "Closed String Tachyons and RG flows",
JHEP 0210 (2002) 051, hep-th/0210054; Takao Suyama,
"Closed String Tachyon Condensation in Supercritical Strings and RG Flows",
hep-th/0510174.}

\nref\horowitz{Gary T. Horowitz, Alan R. Steif, "Singular string
solution with nonsigular initial data", Phys.Lett.B258:91-96,1991.}

\nref\costa{L. Cornalba, Miguel S. Costa, "A New Cosmological Scenario in String Theory",
Phys.Rev. D66 (2002) 066001, hep-th/0203031.}

\nref\niki{Nikita A. Nekrasov, "Milne universe, tachyons and quantum group", Surveys High Energ.Phys.17:115-124,2002,
e-Print Archive: hep-th/0203112.}

\nref\Berkz{Micha Berkooz, Ben Craps, David Kutasov, Govindan Rajesh,
"Comments on cosmological singularities in string theory", JHEP 0303:031,2003,
e-Print Archive: hep-th/0212215.}

\nref\Berka{M. Berkooz, B. Pioline, "Strings in an electric field, and
the Milne Universe", JCAP 0311 (2003) 007, hep-th/0307280.}

\nref\Berkb{M. Berkooz, B. Pioline, M. Rozali, "Closed Strings in
 Misner Space: Cosmological Production of Winding Strings",
JCAP 0408 (2004) 004, hep-th/0405126. }

\nref\Berkc{M. Berkooz, B. Durin, B. Pioline, D. Reichmann,
"Closed Strings in Misner Space: Stringy Fuzziness with a Twist",
 JCAP 0410 (2004) 002, hep-th/0407216.}

\nref\cost{Lorenzo Cornalba, Miguel S. Costa, "Time-dependent Orbifolds and String Cosmology",
Fortsch.Phys. 52 (2004) 145-199, hep-th/0310099.}

\nref\pio{Bruno Durin, Boris Pioline, "Closed strings in Misner
 space: a toy model for a Big Bounce ?", Proceedings of the NATO
 ASI and EC Summer School ``String Theory: from Gauge Interactions
 to Cosmology'', Cargese, France, June 7-19, 2004, hep-th/0501145.}

\nref\Jap{Yasuaki Hikida, Rashmi R. Nayak, Kamal L. Panigrahi,
"D-branes in a Big Bang/Big Crunch Universe: Misner Space", hep-th/0508003.}

\nref\mbb{Ben Craps, Savdeep Sethi, Erik Verlinde, hep-th/0506180;
Miao Li,  hep-th/0506260; Miao Li, Wei Song,  hep-th/0507185;
Sumit R. Das, Jeremy Michelson,  hep-th/0508068; Bin Chen,
hep-th/0508191; Daniel Robbins, Savdeep Sethi, hep-th/0509204;
Micha Berkooz, Zohar Komargodski, Dori Reichmann, Vadim
Shpitalnik, hep-th/0507067; Bin Chen, Ya-li He, Peng Zhang,
hep-th/0509113; Hong-Zhi Chen, Bin Chen, hep-th/0603147; Miao Li,
Wei Song, hep-th/0512335; Takayuki Ishino, Hideo Kodama, Nobuyoshi
Ohta, hep-th/0509173; Rong-Gen Cai, Nobuyoshi Ohta,
hep-th/0601044; Takayuki Ishino, Nobuyoshi Ohta, hep-th/0603215;
Ben Craps, Arvind Rajaraman, Savdeep Sethi, hep-th/0601062; Emil
J. Martinec, Daniel Robbins, Savdeep Sethi, hep-th/0603104.}

\nref\BFSS{T. Banks, W. Fischler, S.H. Shenker, L. Susskind,
"M Theory As A Matrix Model: A Conjecture", Phys.Rev. D55 (1997) 5112-5128, hep-th/9610043.}

\nref\IKKT{N. Ishibashi, H. Kawai, Y. Kitazawa, A. Tsuchiya, "A
Large-N Reduced Model as Superstring",  Nucl.Phys. B498 (1997)
467, hep-th/9612115.}

\nref\seiberg{Nathan Seiberg, "Why is the Matrix Model Correct?",
Phys.Rev.Lett. 79 (1997) 3577-3580, hep-th/9710009; Ashoke Sen,
"D0 Branes on $T^n$ and Matrix Theory ", Adv.Theor.Math.Phys. 2
(1998) 51-59, hep-th/9709220.}

\nref\tseyy{A.A. Tseytlin, "On non-abelian generalisation of
Born-Infeld action in string theory", Nucl.Phys. B501 (1997)
41-52, hep-th/9701125.}

\nref\fdst{Miao Li, "Strings from IIB Matrices",
Nucl.Phys.B499(1997)149-158,
 hep-th/9612222;
 I. Chepelev, Y. Makeenko, K. Zarembo,
"Properties of D-Branes in Matrix Model of IIB Superstring",
Phys.Lett. B400 (1997) 43-51, hep-th/9701151; Ansar Fayyazuddin,
Douglas J. Smith, "P-brane solutions in IKKT IIB matrix theory",
Mod.Phys.Lett. A12 (1997) 1447-1454; hep-th/9701168; A.
Fayyazuddin, Y. Makeenko, P. Olesen, D.J. Smith, K. Zarembo,
"Towards a Non-perturbative Formulation of IIB Superstrings by
Matrix Models", Nucl.Phys. B499 (1997) 159-182, hep-th/9703038.}

\nref\tsey{I. Chepelev, A.A. Tseytlin, "Interactions of type IIB
D-branes from D-instanton matrix model", Nucl.Phys. B511 (1998)
629-646, hep-th/9705120.}

\nref\hag{Miguel S. Costa, Carlos A. R. Herdeiro, J. Penedones, N. Sousa,
 "Hagedorn transition and chronology protection in string theory", Nucl.Phys. B728 (2005) 148-178,
hep-th/0504102.}

\nref\dm{Michael R. Douglas, Gregory Moore, "D-branes, Quivers, and
ALE Instantons", hep-th/9603167. }

\nref\nappi{C. R. Nappi, E. Witten, "A WZW model based on a
non-semi-simple group", Phys. Rev. Lett. 71 (1993) 3751-3753, hep-th/9310112;
 D.I.Olive, E.Rabinovici, A.Schwimmer,
"A Class of String Backgrounds as a Semiclassical Limit of WZW Models", Phys.Lett. B321 (1994) 361-364, hep-th/9311081.}

\nref\kir{Giuseppe D'Appollonio, Elias Kiritsis, "String interactions in
gravitational wave backgrounds", Nucl.Phys. B674 (2003) 80-170, hep-th/0305081;
Yeuk-Kwan E. Cheung, Laurent Freidel, Konstantin Savvidy,
"Strings in Gravimagnetic Fields", JHEP 0402 (2004) 054, hep-th/0309005;
M. Bianchi, G. D'Appollonio, E. Kiritsis, O. Zapata, "String amplitudes in the
Hpp-wave limit of AdS3xS3", JHEP 0404 (2004) 074, hep-th/0402004.}

\nref\kirbr{G. D'Appollonio, E. Kiritsis, "D-branes and BCFT in Hpp-wave backgrounds",
Nucl.Phys. B712 (2005) 433-512, hep-th/0410269.}

\nref\li{Y. Yoneya, in "Wandering in the Fields", eds. K. Kawarabayashi, A. Ukawa (World Scientific, 1987),
p419;  Miao Li, Tamiaki Yoneya, "D-Particle Dynamics and The Space-Time Uncertainty Relation",
Phys.Rev.Lett. 78 (1997) 1219-1222, hep-th/9611072; Miao Li, Tamiaki Yoneya,
"Short-Distance Space-Time Structure and Black Holes in String Theory : A Short Review of the Present Status",
in the special issue of the Journal of Chaos, Solitons and Fractals on "Superstrings, M, F, S...Theory", hep-th/9806240;
 Tamiaki Yoneya, "String Theory and the Space-Time Uncertainty Principle", Prog.Theor.Phys. 103 (2000) 1081-1125,
hep-th/0004074.}

\nref\wij{David Berenstein, Vishnu Jejjala, Robert G. Leigh, "Marginal and Relevant
Deformations of N=4 Field Theories and Non-Commutative Moduli Spaces of Vacua",
Nucl.Phys. B589 (2000) 196, hep-th/0005087; David Berenstein, Vishnu Jejjala, Robert G. Leigh,
"Non-Commutative Moduli Spaces, Dielectric Tori and T-duality", Phys.Lett. B493 (2000) 162,
hep-th/0006168; David Berenstein, Robert G. Leigh, " Resolution of Stringy Singularities by Non-commutative Algebras",
JHEP 0106 (2001) 030, hep-th/0105229; M. Wijnholt, "Parameter Space of Quiver Gauge Theories", hep-th/0512122. }

\nref\huang{Chong-Sun Chu, Brian R. Greene, Gary Shiu,
"Remarks on Inflation and Noncommutative Geometry", Mod.Phys.Lett. A16 (2001) 2231-2240, hep-th/0011241;
Stephon Alexander, Robert Brandenberger, Joao Magueijo, "Non-Commutative Inflation",
Phys.Rev. D67 (2003) 081301, hep-th/0108190; Robert Brandenberger, Pei-Ming Ho,
"Noncommutative Spacetime, Stringy Spacetime Uncertainty Principle, and Density Fluctuations",
Phys.Rev. D66 (2002) 023517; AAPPS Bull. 12N1 (2002) 10-20, hep-th/0203119; Qing Guo Huang, Miao Li,
"CMB Power Spectrum from Noncommutative Spacetime", JHEP 0306 (2003) 014, hep-th/0304203;
Qing-Guo Huang, Miao Li, "Noncommutative Inflation and the CMB Multipoles", JCAP 0311 (2003) 001, astro-ph/0308458;
Qing-Guo Huang, Miao Li, "Power Spectra in Spacetime Noncommutative Inflation",
Nucl.Phys. B713 (2005) 219-234, astro-ph/0311378.}

\nref\btz{Máximo Bañados, Claudio Teitelboim, Jorge Zanelli,
"The Black Hole in Three Dimensional Space Time", Phys.Rev.Lett. 69 (1992) 1849-1851,
hep-th/9204099; Maximo Banados, Marc Henneaux, Claudio Teitelboim, Jorge Zanelli,
"Geometry of the 2+1 Black Hole", Phys.Rev. D48 (1993) 1506-1525, gr-qc/9302012.}

\nref\miao{Miao Li, "Black Holes and Spacetime Physics in String/M Theory",
hep-th/0006024; Miao Li, "Macroscopic Black Holes, Microscopic Black Holes and Noncommutative Membrane",
Class.Quant.Grav. 21 (2004) 3571-3578, hep-th/0311105.}

\nref\zen{Juan M. Romero, J.A. Santiago, J. David Vergara,
"A note about the quantum of area in a non-commutative space",
Phys.Rev. D68 (2003) 067503, hep-th/0305080;
 Axel Krause, "On the Bekenstein-Hawking Entropy, Non-Commutative
 Branes and Logarithmic Corrections", hep-th/0312309; Brian P. Dolan, "Quantum Black Holes: the Event Horizon as a Fuzzy Sphere",
JHEP 0502 (2005) 008, hep-th/0409299; Zen Wei, "Black Hole Complementary Principle
 and The Noncommutative Membrane", hep-th/0511141.}

\newsec{Introduction}

Recently it was found that closed string tachyons lead to many
fascinating phenomena, e.g. smoothing singularities \APS\ \sm\
\horo\ \she\ \tai , changing spacetime topology \sil\ and even
spacetime dimension \hellerman, creating baby universes \sil,
starting and ending time \sm\ \she, and potentially resolving the
black hole information paradox \horo\ \rsimon, many of whom have
never been seen without tachyons (recent progress includes also
\newpaper). It is even more surprising that these can happen at
the level of perturbative string theory, without referring to more
subtle effects. Unfortunately in many situations we can not get
such tachyons where we still expect similar phenomena should
happen. So one
 is led to ask the following
question: can we cook such effects without employing tachyons? From the
world sheet point of view, tachyons are relevant perturbations of
the original CFT. So we are actually asking the question: can
milder (marginal or irrelevant) perturbations be substitutes for such
relevant perturbations? In this note we realize one effect
(resolution of a spacelike singularity) without employing closed
string tachyons. Here winding strings will take the place of
tachyons, providing essentially the same effect.

The resolution of spacelike singularities is one of the most
outstanding problems in the study of quantum gravity.
These singularities make appearance in many black holes and
cosmological models. It is generally very hard to get much
information about them. So in order
to make progress on this issue, more controllable toy models are
proposed, the simplest of which may be the two dimensional Misner
space, which can be defined as the quotient of two dimensional
Minkowski space by a boost transformation.

Misner space can be embedded into string theory by adding 8
additional flat directions, and it is an exact solution of string
theory at least at tree-level \horowitz. The dynamics
of particles and strings in Misner universe were much explored in
the literature (see for example \niki\ \Berkz\ \Berka\ \Berkb\
\Berkc, for comprehensive reviews see \cost\ \pio). In particular, it was
realized in the above papers that winding strings are
pair-produced and they backreact on the geometry. So they are expected play some role in
the resolution of the singularity.
More recently D-brane dynamics in Misner space was also discussed \Jap.

Along another line, in the study of closed string tachyons, Misner space has reemerged as a valuable model\sm. By
imposing anti-periodic boundary conditions for fermions on the
spatial circle, one can get winding tachyons near the singularity
which can significantly deform the original geometry. It is
argued \sm\ that the spacetime near the spacelike singularity will
be replaced by a new phase of the tachyon condensates.

We will use a matrix model to describe such a time-dependent
background, following recent studies\she\ \mbb. One of the
advantages of such holographic description is that backreaction
can be taken into account more naturally, since the geometry and
objects in it are not treated separately as in conventional theory
including perturbative string theory. It was conjectured in \BFSS\
that the large $N$ limit of the supersymmetric matrix quantum
mechanics describing D0-branes provides a holographic description
of M-theory in the light cone frame. In this model, known as BFSS
matrix theory, all spatial dimensions are dynamically generated
while time is put in a prior. Later in \IKKT, another matrix
theory is proposed for IIB theory. This so called IKKT matrix
theory is a 0+0 dimensional theory, in contrast to the 0+1
dimensional BFSS theory. In this theory, both spatial and temporal
dimensions are generated dynamically. And the IKKT action is
essentially just the D-instanton action.

Though lack for a Seiberg-Sen type argument \seiberg\ for
decoupling of other degrees of freedom, there are many checks on
the validity of IKKT matrix model. Fundamental strings and
Dp-branes \IKKT\ \tseyy\ \fdst\ can be constructed from such
matrices, and long-distance interaction potentials of BPS
configurations computed from such matrices match the supergravity
results \tsey.

A Lorentzian orbifolded version of IKKT matrix model, which is
essentially the D-instanton action in Misner space, was recently
proposed in \she\ to describe the full string theory in Misner
space. It was shown there that the Higgs branch of the moduli
space reproduces the original background geometry. And imposing
anti-periodic boundary conditions for fermions, thus cooking some
closed string tachyons near the singularity, the matrix model is
deformed by such tachyon condensates. And in the deformed matrix
model, the original spacelike singularity is shown to be
 replaced by a new phase, where spatial direction does not commute
with the temporal direction. The whole picture in \she\ is consistent with that of \sm. Both
show a breakdown of conventional notions of time near the original singularity.

In this note, we will suggest that the pair-produced winding
strings can play the same role as the closed string tachyons in
\sm\ and \she, namely the condensation of these winding strings
will also smooth the spacelike singularity. Interestingly in a
recent paper \hag, it was found that winding string condensation
can also restore chronology in some tricky backgrounds.

Our result is new in the following two respects.  First, from the
line of study of closed string tachyons \sm\ \she\ \tai, we
generalize the whole picture there to more generic situations,
without the previous need to cook up the particular fermion
boundary conditions. Our result further indicats that the
fascinating effects encountered within closed string tachyons may
well exist in other cases without closed string tachyons. Second,
viewed from the line of study of winding string effects in Misner
space \Berka\ \Berkb\ \Berkc, we step into the open string sector
and further propose a non-perturbative description. Near the
singularity the open string description has some superiority to
the closed string description. We show that near the singularity,
the conventional notion of space-time breaks down; nonetheless
there exists a matrix model description in the open string sector.
And in particular we show that space and time turn out to be
noncommutative near the singularity.

There is one point deserving further clarification. One may think
that winding strings can not cure the breakdown of perturbative
string theory near cosmological singularities, since the breakdown
already happens at tree level \Berkz, while winding effects are
higher order. Here we need to discriminate between the effects of
a few winding strings and the effects of condensates of winding
strings. The former is demonstrated in \Berkc\ by calculating
amplitudes involving winding strings and such effects are
certainly higher order. However for the case of condensates of
winding strings where backreactions greatly change the background,
 the effects are lowest order. In fact, such effects are encoded in disk diagrams with the
insertion of the winding strings at the center and are thus
independent of string coupling, or of order $g_s^0$ . Such effects
call for a nonvanishing vacuum expectation value of the winding
string field, thus condensation of winding strings. And as we will
review in this note, there does exist a condensation of winding
strings with zero boost momentum near the cosmological singularity
\Berka. We note that perturbative string theory, in expansion of
positive powers of string coupling, can not describe such effects
in principle.

The layout of this note is as follows. In section 2, we review
some aspects of the Misner geometry, particularly cosmological
production and condensation of winding strings. Then in section 3,
we encapsulate the backreaction of these winding strings in the
deformed D-instanton matrix model, and read from the new moduli
space the fate of the spacelike singularity. In section 4, we
study the disk and cylinder amplitudes which are crucial for our
matrix model. Our conclusions are presented in section 5, together
with some comments on open issues.

\newsec{Winding string production in Misner space}
In this section we review mainly Berkooz et al.'s work \Berka\
\Berkb\ showing winding string condensation in Misner space, using
bosonic string as an example.

Misner space is an orbifold of 1+1-dimensional Minkowski space
\eqn\mis{ds^2=-2dx^+dx^-}by the identification \eqn\orb{x^+\sim
e^{2\pi\gamma}x^+,\quad x^-\sim e^{-2\pi\gamma}x^- .} Coordinate
transformation \eqn\tran{x^+={T\over \sqrt{2}}
e^{\gamma\theta},\quad x^-={T\over \sqrt{2}} e^{-\gamma\theta}}can
be made to write Misner space as
\eqn\misa{ds^2=-dT^2+\gamma^2T^2d\theta^2,}with
$\theta\cong\theta+2\pi.$ It is easy to see from \misa\ that this
space-time contains two cosmological regions connected by a
space-like singularity.

There are generally two kinds of closed strings in Misner universe: twisted and untwisted.
Untwisted states include in particular the gravitons and their behaviors are particle-like. Their
wave functions can be obtained by superposing a plane wave in the parent Minkowski space
 with its images under the boost \orb, and is written as \niki\ \Berkb\
\eqn\untw{f_{j,m^2,s}(x^+,x^-)=\int^{\infty}_{-\infty}dv e^{ip^+X^-e^{-2\pi\gamma v}+ip^-X^+e^{2\pi\gamma v}+ivj+vs},}
with $j$ the boost momentum, $m$ the mass, and $s$ the $SO(1,1)$ spin in $R^{(1,1)}$.

Due to the orbifold projection \orb, new twisted sectors arise in
Misner space with strings satisfying
\eqn\twist{X^{\pm}(\tau,\sigma+2\pi)=e^{\pm 2\pi\gamma
w}X^{\pm}(\tau,\sigma),} where the winding number $w$ is an
integer. Many mysterious of the Misner universe have origin from
these winding strings. It was shown in \Berka\ that there exists a
delta-function normalizable continuum of physical twisted states,
which can be pair produced in analogy with the Schwinger effect in
an electric field \Berka\ \Berkb.

Such production can be seen most straightly in closed string
one-loop vacuum amplitude. There are some subtleties in the
world-sheet and spacetime analytic continuation, and it was
checked in \Berkb\ that a well-defined amplitude should be an
integral with Euclidean world-sheet and Minkowskian spacetime and
of the form \eqn\abos{A=\int_F
\sum_{l,w=-\infty}^{\infty}{{d\rho_1
d\rho_2}\over{{(2\pi^2\rho_2)}^{13}}}{{e^{-2\pi\gamma^2w^2\rho_2-
{{R^2}\over{4\pi\rho_2}}\mid
l+w\rho\mid^2}}\over{\mid\eta^{21}(\rho)\theta_1[i\gamma(l+w\rho);\rho]\mid^2}},}with
$F$ the fundamental domain of the moduli space of the torus,
parameterized by $\rho=\rho_1+i\rho_2$, and $\theta_1$ and $\eta$
are the Jacobi and Dedekind functions. The amplitude has poles at
\eqn\poles{\rho={{k-i\gamma l}\over{n+i\gamma w}},}with integer
$k$ parameterizing the poles and integer $n$ labelling the
vanishing factor in the Jacobi theta function. We are interested
in the logarithm of the one-loop amplitude $A$, which is known as
free energy, since its imaginary part gives the total decay rate
of the vacuum. For the real worldsheet action, nonvanishing
imaginary part can only arise from the above singularities \poles.

To explore the spacetime interpretation of the amplitude,
especially in the case with non-vanishing winding number, analogy
with charged particles in the electric field was drawn \Berka\
\Berkb. To use the previously developed semi-classical methods,
they write the amplitude as the closed string path integral
\eqn\aint{\ln A=\int_F{{d\rho_1 d\rho_2}\over
{\rho_2}}\sum_{l,w=0}^{\infty}\int[DX]e^{-S_{l,w}}}with action
\eqn\wact{S={1\over2}\int_0^{2\pi\rho_2}d\tau
\int_0^{2\pi}d\sigma(\p_{\tau}X^+\p_{\tau}X^-
+\p_{\sigma}X^+\p_{\sigma}X^-) ,}where we integrate over paths
restricted to periodic configurations (with string winding number
$w$)
\eqn\ppath{\eqalign{X^{\pm}(\sigma+2\pi,\tau)&=e^{\pm2\pi\gamma
w}X^{\pm}(\sigma,\tau) \cr
X^{\pm}(\sigma+2\pi\rho_1,\tau+2\pi\rho_2)&=e^{\mp2\pi\gamma
l}X^{\pm}(\sigma,\tau)}.} Truncate to quasi-zero-modes
\eqn\qzero{X^{\pm}=\pm{1\over{2\nu}}\alpha^\pm
e^{\mp(\nu\sigma-iA\tau)}\mp{1\over{2\nu}}{\tilde\alpha}^\pm
e^{\mp(\nu\sigma+iA\tau)},}with $\nu=-\gamma w$ and
\eqn\aatilde{A={k\over{\rho_2}}-i\gamma{{l+\rho_1
w}\over{\rho_2}}, \quad {\tilde A}={{\tilde
k}\over{\rho_2}}+i\gamma{{l+\rho_1 w}\over{\rho_2}},}and
$k,{\tilde k}$ labelling the periodic trajectory, and
\eqn\alalt{\alpha^{\pm}=\pm Re^{\pm\eta}/{\sqrt 2},\quad
{\tilde\alpha}^{\pm}=\pm {\tilde R}e^{\pm{\tilde\eta}}/{\sqrt 2}.}
In fact, \ppath\ has the interpretation as an Euclidean worldsheet
interpolating between two Lorentzian worldsheets which correspond
to winding strings propagating in Misner space. And later it will
account for the tunnelling effect for winding string pair
production.

Now the amplitude becomes an integral of the form \eqn\fint{\int
d\rho_1 d\rho_2 \int dR d{\tilde R}\int d\eta d{\tilde\eta}
e^{-S(\rho_1,\rho_2;R,{\tilde R};\eta,{\tilde\eta})},} where the
integration over $\eta,{\tilde\eta}$ can be easily performed, but
integration over $R,{\tilde R}$ leads to divergences. The trick is
to integrate out $\rho$ first. The Gaussian integral with respect
to $\rho_1$ is dominated by a saddle point at
\eqn\sadrho{\rho_1=-{{\gamma l}\over \nu}+i{{{\tilde k}{\tilde
R}^2-kR^2}\over{\nu(R^2+{\tilde
R}^2)}}-2i{{j\nu\rho_2}\over{R^2+{\tilde R}^2}}.}

The crucial point here is that this saddle point is a local
maximum of the Euclidean action. Perturbations around this point
lead to instabilities, indicating spontaneous pair production of
winding strings by condensing the $\rho_1$
modes.$^*$\footnote{}{*We should also remind the reader that the
imaginary part of the free energy, and thus the total pair
production rate (with observing time infinitely long), vanishes,
which is not surprising since the globally defined $in$ and $out$
vacua are identical for Misner space, though there are pair
productions near the singularity \Berkb.} Similar instabilities
were found in \she\ by perturbing the matrix model action there.

Winding string pair production can also be seen by directly quantizing strings in Misner space \Berka\ \Berkb,
taking into account effects from Rindler regions of the extended Misner space.
Excited modes of the bosonic part and fermionic modes, both zero- and excited, can
be quantized using the same methods as in flat space. Subtleties lie in the bosonic zero-modes. In fact the string center
of mass moves in an inverted harmonic potential, where similarities to charged particles in an electric field
will also be found. The wave function in one of the Rindler patches
 is the zero-energy modes of the Schrodinger operator $-{d^2\over{dy^2}}+V(y)$,
with the potential \eqn\poten{V(y)={{M^2{\tilde
M}^2}\over{\nu^2}}-({{{M^2+{\tilde
M}^2}\over{2\nu}}-{\nu\over2}r^2})^2,}where $r=e^y$ and
\eqn\mass{M^2=\al^+_0\al^-_0+\al^-_0\al^+_0, \quad {\tilde
M}^2=\at^+_0\at^-_0+\at^-_0\at^+_0} with string zero modes
$\al^{\pm}_0$ and $\at^{\pm}_0$, coming from the Virasoro
conditions. Tunnelling events under the above potential barrier
correspond to local production of winding strings in Misner space,
together with annihilation of long strings in the Rindler patch.
Such induced string production gives complementary evidence for
the spontaneous string production above.

By calculating the Bogolubov coefficients for the overlap of the
wave functions, the induced pair production rate, i.e. the
transition coefficient in Misner region, can be shown to be \Berka
\eqn\qqq{q=e^{-\pi M^2 /2\nu}{{\cosh(\pi {\tilde
M}^2/2\nu)}\over{|\sinh\pi j|}},} which, near the singularity
where winding strings can be very light,
 is of order unity with large winding number $w$.
  And thus strings with large winding number can be copiously
produced.

A semi-classical estimation of the number of string pairs produced
is carried out in \Berka\ with the assumption that the
distribution is homogeneous in the $(x_0^+,x^-_0)$ phase space,
with the result \eqn\numpp{p(X^+,X^-)={1\over{\sqrt{(j-\nu
X^+X^-)^2+2M^2X^+X^-}}}.} With vanishing boost momentum $j$, the
production number diverges at the singularity where
$T^2=2X^+X^-=0$, leading to condensation of winding strings here.
And in the infinite past and infinite future, where $T^2$ goes to
infinity, there is no string production and we recover the flat
space physics correctly.

\newsec{Winding-string-deformed matrix model for Misner space}
Now we go on to ask: what is the effect of the condensation of the pair produced winding
strings to the background geometry? It is generally difficult to encapsulate these backreactions.
In the perturbative string formalism, Berkooz et al calculated \Berkc\ the effects of these
winding strings to scattering amplitudes involving
both twisted and untwisted strings, where induced fuzziness, much larger than string scale, was found.
 Here we will employ the recently proposed nonperturbative formulation
of string theory in Misner space via matrix model \she. Matrix models have the advantage of treating
the geometry and objects in it more democratically where all are matrices, and hence the backreaction problem
 can potentially be
better formulated.

The matrix model action in \she\ is taken to be the D-instanton action in Misner space
which thus can be calculated using the now standard methods for D-branes on orbifolds. With the coordinates $x^{\pm}$
promoted to matrices $X^{\pm}$, the action
can be written in the following form
\eqn\acta{{\eqalign{S&=-{1\over
g^2}\int_0^{2\pi}{{d\sigma}\over{2\pi}}\Tr
\biggm([e^{i2\pi\gamma{d\over{d\sigma}}}X^+(\sigma)]X^-(\sigma)-
[e^{-i2\pi\gamma{d\over{d\sigma}}}X^-(\sigma)]X^+(\sigma)\biggm)^2\cr
&=-{1\over
g^2}\int_0^{2\pi}{{d\sigma}\over{2\pi}}\Tr\biggm[X^+(\sigma+i2\pi\gamma)X^-(\sigma)
-X^-(\sigma-i2\pi\gamma)X^+(\sigma)\biggm]^2},}} which is complemented by
the symmetry \eqn\sym{\eqalign{X^+(\sigma)&\rightarrow
e^{2\pi\gamma}X^+(\sigma),\cr X^-(\sigma)&\rightarrow
e^{-2\pi\gamma}X^-(\sigma),}}inherited from the orbifold
projection \orb. The integral above comes from Fourier transformation of the infinite summation
of the winding modes. The Higgs branch of the matrix model is shown to reproduce the original
Misner space.

The key point to resolve the spacelike singularity is that, when winding
strings condense in Misner space, the matrix model \acta\
 gets deformed
\eqn\actde{S=-{1\over
g^2}\int_0^{2\pi}{{d\sigma}\over{2\pi}}\biggm[X^+(\sigma+i2\pi\gamma)X^-(\sigma)
-X^-(\sigma-i2\pi\gamma)X^+(\sigma)-U(\sigma)\biggm]^2,}
where the $U(\sigma)$ term is the vacuum expectation value of the twisted fields. This can be seen
as follows \she.

 It was shown by Douglas and Moore in
  \dm\ that the leading effect of twisted fields on the Euclidean orbifolds is to
induce a FI-type term in the D-brane potential. This effect comes
from the disk amplitude $<\phi_k XX>$ with one insertion of the twisted sector
field $\phi_k$ at the center and two open string vertex operators at
the boundary. With a detailed analysis of the full quiver gauge
theory, which provides a description for D-branes on the orbifolds, they combine the
FI term with the Born-Infeld action and the kinetic energies of the
hypermultiplets, and then integrate out the auxiliary D-fields in
the vectormultiplet, to find that the effect of the twisted sector
fields is to add a term in the complete square. In our case, we are
dealing with a Lorentzian orbifold which is more subtle than its Euclidean cousin.
However to study the D-instanton theory,
a Wick rotation can be performed to go to the Euclidean case, where the result of \dm\ can be consulted to
 finally get the deformed matrix model \actde.

It will also be good to perform an explicit computation of the
disk amplitude $<\phi_k XX>$ in Misner space. Although we still do
not know how to write out the twisted vertex operators, some
techniques have been developed in \Berkc\ to calculate such
scattering amplitudes. There they observed that the vertex
operators and amplitudes in Misner space can be analytically
continued to corresponding objects in Nappi-Witten plane wave
background, where analogous problems have already been solved,
taking advantage of the fact that it is a Wess-Zumino-Witten
model\nappi. Particularly they show that the two-point,
three-point and four-point functions of twisted states in Misner
space match the plane wave result \kir. We note that the above
amplitude $<\phi_k XX>$ has also been calculated in the plane wave
background \kirbr\ and is non-vanishing, providing another check
for the FI-type coupling used above.

In some sector of the full theory (with anti-periodic boundary
conditions for fermions on the $\theta$ circle), closed tachyon
condensations can cook such vacuum expectation values for the
twisted fields \sm\ \she. In Misner space, generally there are no
closed string tachyons. But we can have other sources for such
twisted field vacuum expectation values.
 In the above section we show that winding strings can be created and their condensation will also
play such a role.

With such a deformed matrix model, we can show that the spacelike singularity gets
non-commutative deformation, following the same process in \she. The vacua are
deformed to \eqn\vac{[X^+(\sigma),X^-(\sigma)]=U(\sigma),} where we have used the approximation $\gamma=0$, since
the winding string condensates encoded in $U(\sigma)$ now dominate.

From \vac\ we see that the spacelike singularity in Misner space
gets space-time noncommutative deformation \li.$^{**}$
\footnote{}{$**$Non-commutative deformations of Calabi-Yau
singularities have been studied in \wij. Cosmological consequences
of space-time noncommutativity were discussed in \huang.} Far from
the singularity,
 in the infinite past and infinite future, there is no winding string production
(total number $p$ goes to zero as $T^2=2X^+X^-$ goes to $\infty$
in \numpp), we recover the results in flat space, where we have a
well defined picture for conventional geometry. Near the
singularity, there are plenty of pair productions ($p$ large or
even diverging), ruining conventional notion of geometry. Think of
an observer starting in the infinite future where she has a good
notion of time, then let her go back in time. With the physical
data at hand and
 assuming legality
of conventional spacetime geometry, she may deduce that she "will" encounter a spacelike
singularity. But "before" that happens, at some time scale $t_c$,
she will find that she
has already lost the notion of "time". Thus time itself is created at some "time" in the history. This observer
resembles so much human beings today, and our universe may have gone through
similar stages in its early days.

 Correspondingly the observers starting from infinite past will find at some time scale $t_c$, the notion of time ends.
Thus Misner space provides a concrete example for starting and ending time in string theory \sm\ \she.

It is interesting to ask about the noncommutative scale $t_c$. Winding strings induce in the untwisted closed
string vertex operator a factor $e^{-k^+k^-\Delta(\nu)}$, leading to fuzziness
on a scale $\sqrt {\ap \log w}$, with $w$ the winding number. However D-instantons are point-like
objects, and open strings attached to them have no propagating momentum and no such fuzziness.

\newsec{Bulk boundary coupling}
One particular feature of our matrix model is that the backreaction of
the winding strings is encoded in the bulk boundary coupling. So it is tempting to look more
closely into such disk amplitudes e.g. $<\phi_k XX>$. Unfortunately we still do not
know how to deal with twisted vertex operators with a continuous
spectrum. However, since we are interested in the average effect
of all the winding strings condensed, this problem can be overcome by using
the optical theorem \eqn\opt{2Im
<V_2V_1V_1V_2>_{1-loop}=\sum_{k,N,\tilde{N}} |<\phi_k
V_1V_2>_{disk}|^2,}where the sum also includes other quantum
numbers of the closed string inserted in the interior the disk. We
will go on to calculate the four-open-string cylinder amplitude,
which contains information about the effect of the condensed
winding strings. One can also study the case in a particular
winding sector, summing over the excitations only, that is to
calculate\eqn\opt{2Im
<V_2V_1V_1V_2>^{k}_{1-loop}=\sum_{N,\tilde{N}} |<\phi_k
V_1V_2>_{disk}|^2,}

Open strings on the D-instanton satisfy Dirichlet boundary
conditions in all space-time dimensions. So there are so physical
propagating open strings. To see the coupling of the D-instanton
to the winding strings, one can study the simplest case with only
open string tachyons, whose vertex operators do not have momentum
dependence. We consider 26 dimensional bosonic string theory with
the D-instantons also localized in the extra dimensions, and the
one-loop amplitude can be deduced from the Misner part partition
function \tai\ by adding extra 24 dimensions and the ghost part
\eqn\zpar{<V_TV_TV_TV_T>_{1-loop}=\sum^{\infty}_{k=-\infty}\int
{{dt}\over t}\xi^{k(m-m')}
{{2i\sinh(\pi|k|\gamma)}\over{\theta_1(i|k|\gamma|it)\eta(it)^{21}}},}
where $m,m'$ are Chan-Paton indices, and $\xi$ is a complex number
introduced to ensure that Chan-Paton indices are invariant under
the orbifold action. Under modular transformation $t\rightarrow
s=1/t$, the above expression can be written in the closed string
channel as
\eqn\zs{<V_TV_TV_TV_T>_{1-loop}=\sum^{\infty}_{k=-\infty}\int
ds\xi^{k(m-m')} {{2\sinh(\pi|k|\gamma)e^{-\pi
sk^2\gamma^2}}\over{\theta_1(|k|\gamma s|is)\eta(is)^{21}}}.}

 The imaginary part of \zs\ can be more easily calculated by
 picking up poles of the theta function at $s={n\over{|k|\gamma}}$
 with $n=1,2,...$, and the result for a particular sector with winding number
 $k$ is \eqn\wtt{\sum_{N,\tilde{N}}|<\phi_k T
T>_{disk}|^2=\xi^{k(m-m')}
\sum^{\infty}_{n=1}(-1)^{n+1}2\sinh(\pi|k|\gamma)\sum_{states}e^{-2\pi
n{{N-1}\over{|k|\gamma}}-\pi n|k|\gamma},}where we follow the
notation of \Jap\ to define
\eqn\eet{\eta(is)^{-24}=\sum_{states}e^{-2\pi s(N-1)}.} \wtt\
diverges when $n<|k|\gamma$, due to the Hagedorn behavior
$\rho_N\sim e^{2\pi N}$.

To illustrate more clearly the closed string channel
interpretation of the one-loop amplitude, one can rewrite \zs\ as
\eqn\zss{\eqalign{<V_TV_TV_TV_T>=\sum^{\infty}_{k=-\infty}&\xi^{k(m-m')}{\ap\over{2\pi}}\int
ds\int d(w^2)\sum_{states} \cr &C \rho(w^2)e^{-2\pi
s(w^2|k|\gamma+N-1+\half
k^2\gamma^2)},}}or\eqn\zsss{<V_TV_TV_TV_T>=\sum^{\infty}_{k=-\infty}\xi^{k(m-m')}{\ap\over{2\pi}}\int
d(w^2)\sum_{states} {{C \rho(w^2)}\over{2\pi
(w^2|k|\gamma+N-1+\half k^2\gamma^2)}},}where $C$ is the
normalization coefficient, and $\rho(w^2)=\half(1+\tanh\pi w^2)$.
Thus we see that the poles come from the closed string propagator
$L_0+\tilde{L}_0=2$.

In the study of closed string scattering amplitudes in \Berkc, it
was found that there exists a new scale $\sqrt{\ap \log w}$, with
$w$ the string winding number. This scale characterizes the
fuzziness caused by winding strings. Since $w$ can be large, this
scale can be much larger than string scale. It is interesting to ask
whether we can see this scale in our matrix model, particularly from
the disk amplitudes calculated above. Unfortunately the answer is
negative.

In the closed string amplitudes, for example the spherical amplitude
with two twisted strings and one untwisted string, there is always a
form factor of the form $e^{-p^+p^-\Delta(\nu)}$, with $p$ the
momentum. This comes from the expectation value of the excitation
part of the un-twisted string vortex operator in the background of
twisted strings. It tells us that winding strings polarize untwisted
strings into a cloud of rms size $\sqrt{\ap\Delta(\nu)}$ \Berkc. At
large winding number, $\Delta(\nu)$ scales as $\log\nu$.

In the above D-instanton disk amplitudes, the situation is
different. D-instantons have the peculiar property that they are
totally localized in spacetime. And the open strings representing
their excitations have no propagating degrees of freedom and thus
are not polarized by the winding strings. Technically, one can see
that except for an overall dependence on $\ap$ justifying their
dimensions, the above D-instanton amplitudes are only functions of
the dimensionless parameter $\nu=|k|\gamma$. No new scale is
generated as in the closed string scattering amplitudes.

If we view our matrix model as a microscopic description of Misner
space, and the geometry as well as closed strings as derived
concepts, it seems that we also need to regard the new scale in
closed string scattering amplitudes as an emergent phenomena.

\newsec{Discussions }
The resolution of spacelike singularity stands out long as a great
challenge in the understanding of quantum theory of gravity. In
this note, following our previous work in \she, building on the
series of elegant work by Nekrasov \niki\ and Berkooz et al.
\Berkz\ \Berka\ \Berkb\ \Berkc, we make modest progress in this
direction. Employing the tool of matrix model, we show that due to
winding string pair production, the original singular geometry is
not a suitable description of the vacuum. Near the region of the
assumed-to-be singularity, the notion of conventional spacetime
breaks down. Nevertheless the system still has a mathematical
description, which is now a space-time noncommutative geometry.

The singularity resolution mechanism in this note has some
interesting characteristics. First, the resolution is stringy. We
have to employ stringy degrees of freedom, i.e. winding strings,
which will be unseen at the level of field theory. The equation of
state of winding strings is very different from ordinary
particles. They provide an effective cosmological constant for
this two dimensional spacetime, and thus hold on the big crunch.
Second, the resolution of this singularity is necessarily quantum
mechanical (in contrast to the case in \she). From the worldsheet
point of view, tachyons give relevant perturbations to the
original CFT describing the original singular spacetime, driving
it to another theory, while these winding strings give generally
irrelevant perturbations, whose effects at first glance will be
negligible. But we see above that through quantum creation and
subsequently backreacting on the geometry, they also change the
background remarkably. The mechanism is also robust. It requires
only the existence of a deformation term with no further request
of its detailed form.

There are still many interesting open questions regarding Misner space.
For example, how to calculate the winding string production rate in the deformed background where
the divergent result
in the original spacetime is expected to be curable.
Some attempts have been made in \Berkb. And it is tempting to speculate that
 the pair production spectrum may be thermal similar to
amplitudes in the tachyon condensate phase calculated in \sm.

Since non-rotating BTZ black hole with $M>0$ \btz\ also possesses a Misner type spacelike singularity,
it is interesting to find out whether the singularity resolution mechanism in Misner space can shed some light on issues there.
Space-time noncommutativity on stretched horizon of black holes was observed in \miao\ (see also \zen).

{\bf Acknowledgments}

We thank W. He, M. Li and W. S. Xu for valuable discussions and
insightful comments.

\listrefs
\end